\documentclass[final,
              3p,
							times,
							twocolumn
							]{elsarticle}

\usepackage[utf8]{inputenc} %deutsche Zeichen
\usepackage{graphicx}
\usepackage{xcolor}
\usepackage{slashed}
\usepackage{amsmath}
\usepackage{microtype}

\def\nin{\noindent}

\journal{Nuc. Phys. (Proc. Suppl.)}

\begin{document}

\begin{frontmatter}

\title{Defining medium-specific condensates in QCD sum rules for \(D\) and \(B\) mesons}

 \author[label1]{Thomas Buchheim\corref{cor1}}
  \address[label1]{Helmholtz-Zentrum Dresden-Rossendorf, PF 510119, D-01314 Dresden, Germany}
\cortext[cor1]{Speaker}
\ead{t.buchheim@hzdr.de}

\author[label2]{Thomas Hilger}
  \address[label2]{University of Graz, Institute of Physics, NAWI Graz, A-8010 Graz, Austria}
\ead{thomas.hilger@uni-graz.at}

 \author[label1,label3]{Burkhard K\"{a}mpfer}
  \address[label3]{Technische Universit\"{a}t Dresden, Institut f\"{u}r Theoretische Physik, D-01062 Dresden, Germany}
\ead{b.kaempfer@hzdr.de}

\begin{abstract}
\noindent
QCD sum rules for \(D\) and \(B\) mesons in a strongly interacting environment require the decomposition of Gibbs averaged operators related to tensor structures. We present a procedure to decompose these operators into vacuum and medium-specific parts, thus defining plain medium-specific condensates with coefficients vanishing in vacuum. Our decomposition allows for an identification of potential elements of order parameters for chiral restoration, in particular for higher mass-dimension quark-condensates which would be masked otherwise if operators with uncontracted Lorentz indices would be linked to DIS amplitudes. 

\end{abstract}


\end{frontmatter}

% \linenumbers

\section{Introduction}
\label{sec:intro}
\nin
QCD sum rules \cite{svz79,narison} represent a useful tool to get access to hadron properties without cumbersome lattice evaluations. Following Wilson's operator product expansion (OPE), a current-current operator is decomposed into a series of basic operators and Wilson coefficients thus separating short-range and long-range phenomena. The expectation values of the basis operators are called condensates, thought to be universal, and in \cite{brodsky10} even assumed to be intimately related to hadron wave functions. According to \cite{hatsukoike,furnstahl}, the condensates change in a strongly interacting environment. Since the condensates are related to hadron spectral functions the properties of low mass hadrons are therefore closely connected with vacuum structure. Hence changes of hadron properties in medium carry signals of the way the vacuum changes in a nuclear environment \cite{weise94}.

The next-generation experiments at FAIR are going to address the properties of charm (bottom) mesons in nuclear matter \cite{CBMBook}, among other important issues. Motivated by this perspective we elaborate here an important element of in-medium QCD sum rules for \(D\) and \(B\) mesons (i.\,e.\ colorfree composites of a heavy and a light valence quark), especially for higher-mass dimension condensates, e.\,g.\ four-quark condensates, which can potentially serve for constructing order parameters for chiral restoration.

\section{Operator product expansion}
\label{OPE}
\nin
The causal current-current correlator reads generally
\begin{align}
	\Pi(q) = i \int d^4x e^{iqx} \, \langle\!\langle \mathrm{T}\left[j(x) j^\dagger (0)\right] \rangle\!\rangle_{T,\,\mu} \, ,
\end{align}
where \(\langle\!\langle \ldots \rangle\!\rangle_{T,\,\mu}\) means Gibbs averaging with Lagrange parameters \(T\) (temperature) and \(\mu\) (baryo-chemical potential). The operators \(j\) reflect the quantum numbers of the hadron under consideration. In vacuum, \(T = 0\), \(\mu = 0\), the Gibbs average reduces to the vacuum expectation value. Note that the requirement of a continuous transition to vacuum imposes constraints.

In a medium, which is supposed to be in (local) equilibrium, the Lorentz covariance of the vacuum is broken. In practice this implies a tensor decomposition with the elements \(g_{\mu\nu}\) (metric tensor), \(\varepsilon_{\mu\nu\lambda\sigma}\) (Levi-Civita symbol) and \(v_\mu\) (medium four-velocity), at least. Our goal here is an easy formalism for identifying terms, both scalar coefficients and tensor structures, being specific for the presence of an ambient medium. In doing so we arrive at structures with well defined vacuum expressions in "turning off the medium".

In particular, we decompose quark expectation values with a certain number of uncontracted Lorentz indices to obtain only scalar quark condensates as coefficients. This enables us to identify elements qualifying for order parameters of chiral restoration. These would be masked if one directly links the operators with uncontracted Lorentz indices (cf. higher-twist operators) to DIS amplitudes as often done (c.f.\ \cite{leupold98}).

\section{Decomposition of Lorentz structures}
\label{sec:ProjLInd}
\nin
Executing the OPE gives Gibbs averaged structures \(\langle O_{\vec{\mu}_n} \rangle\), where \(\vec{\mu}_n\) is a short hand notation for a sequence of \(n\) Lorentz indices and \(\langle \ldots \rangle\) is an abbreviation of \(\langle\!\langle \ldots \rangle\!\rangle_{\text{T},\,\mu}\). Our goal is the suitable decomposition of the Gibbs average. The four steps to do so are:

(\(i\)) In vacuum, a complete tensor basis \(\overset{\text{vac}}{p}{}_{\vec{\mu}_n}^j\) has \(j_\text{max}\) elements built up by \(g_{\mu\nu}\) and \(\varepsilon_{\mu\nu\lambda\sigma}\) mentioned above yielding a decomposition
\begin{align}
	\langle O_{\vec{\mu}_n} \rangle & = \overset{\text{vac}}{a}{}_j \overset{\text{vac}}{p}{}_{\vec{\mu}_n}^j \, ,
	\label{eq:vacansatz}
\end{align}
and the projection
\begin{align}
	\overset{\text{vac}}{c}{}^{j'} & = \langle O_{\vec{\mu}_n} \rangle\; \overset{\text{vac}}{p}{}^{\vec{\mu}_n \, j'} \nonumber\\
	& = \overset{\text{vac}}{a}{}_j \overset{\text{vac}}{p}{}_{\vec{\mu}_n}^j \overset{\text{vac}}{p}{}^{\vec{\mu}_n \, j'} = \overset{\text{vac}}{a}{}_j \overset{\text{vac}}{\mathbf{P}}{}_n^{jj'} \, .
	\label{eq:vacproj}
\end{align}
Since the vacuum projection matrix \(\overset{\text{vac}}{\mathbf{P}}{}_n^{jj'}\) is symmetric and invertible one obtains the eventual expression of the decomposition coefficients
\begin{align}
	\overset{\text{vac}}{a}{}_j & = \left( \overset{\text{vac}}{\mathbf{P}}{}_n \right)^{-1}_{jj'} \overset{\text{vac}}{c}{}^{j'} \, .
	\label{eq:vacresult}
\end{align}

(\(ii\)) In medium, the complete tensor basis is extended by new elements, e.\,g., incorporating the medium's four-velocity \(v_\mu\) forming \(P^i_{\vec{\mu}_n}\) with \(i_\text{max} > j_\text{max}\), where the \(j_\text{max}\) vacuum structures are noted first (cf.\ Eq.~\eqref{eq:projvec4LInd} exemplarily). Thus, \(P^i_{\vec{\mu}_n}\) provides the decomposition
\begin{align}
	\langle O_{\vec{\mu}_n} \rangle = A_i P^i_{\vec{\mu}_n} \, ,
	\label{eq:ansatzfullmed}
\end{align}
which can be split into a medium-specific part and the above defined vacuum part:
\begin{align}
	\langle O_{\vec{\mu}_n} \rangle & = \overset{\text{vac}}{a}{}_j \overset{\text{vac}}{p}{}_{\vec{\mu}_n}^j +  \overset{\text{med}}{a}{}_{k} \overset{\text{med}}{p}{}_{\vec{\mu}_n}^{k}
	\label{eq:ansatzspli}
\end{align}
with
\begin{align}
	\overset{\text{med}}{a}{}_{k} \overset{\text{med}}{p}{}_{\vec{\mu}_n}^{k} & =  A_i P^i_{\vec{\mu}_n} - \overset{\text{vac}}{a}{}_{j} \overset{\text{vac}}{p}{}_{\vec{\mu}_n}^{j} \, ,
	\label{eq:medansatz}
\end{align}
i.\,e.\ when turning back to vacuum \(\overset{\text{med}}{a}{}_{k} \overset{\text{med}}{p}{}_{\vec{\mu}_n}^{k} = 0\). Since \(\overset{\text{med}}{p}{}_{\vec{\mu}_n}^{k}\) is not known a priori and the splitting \eqref{eq:ansatzspli} remains rather formal.
However, the components \(A_i\) can be identified with combinations of scalar condensates. Details of determining the \(A_i\) by projection are provided in \cite{buchheim14} for structures with up to five Lorentz indices. A disadvantage of the structure \(P^i_{\vec{\mu}_n}\) in \eqref{eq:ansatzfullmed} derived in \cite{buchheim14} is, however, the uncomfortable mixing of vacuum and medium-specific terms occurring for \(n = \text{even}\).

(\(iii\)) Therefor, we here propose an alternative approach based on the requirement
\begin{align}
	0 & = \overset{\text{med}}{p}{}_{\vec{\mu}_n}^{k} \; \overset{\text{vac}}{p}{}^{\vec{\mu}_n\, j}
	\label{eq:ortho}
\end{align}
guaranteeing the desired splitting into vacuum and medium-specific parts. Both approaches give the same decomposition results, i.\,e.\ Eq.~\eqref{eq:medansatz} is reproduced correctly for an arbitrary number of Lorentz indices. Eq.~\eqref{eq:ortho} implies
\begin{align}
	0	& = \left(  A_i P^i_{\vec{\mu}_n} - \overset{\text{vac}}{a}{}_{j'} \overset{\text{vac}}{p}{}_{\vec{\mu}_n}^{j'} \right) \overset{\text{vac}}{p}{}^{{\vec{\mu}_n}\, j} \, ,
\end{align}
which facilitates \(j_\text{max}\) constraints on the coefficients \(A_i\) and \(\overset{\text{vac}}{a}{}_j\), i.\,e.\ obviously only \(i_\text{max}\) of them are independent. Substituting the obtained constraints% \(A_j = f(A_{ i\neq j}, \overset{\text{vac}}{a}{}_{j'})\)
\begin{align}
	A_j = \overset{\text{vac}}{a}_j - \sum_{i = 1}^{i_\text{max} - j_\text{max}} A_{i + j_\text{max}} P_{\vec{\mu}_n}^{i + j_\text{max}} \overset{\text{vac}}{p}{}^{\vec{\mu}_n \, j'} \left( \overset{\text{vac}}{\mathbf{P}}_n \right)^{-1}_{jj'}
\end{align}
in Eq.~\eqref{eq:ansatzfullmed} allows for an identification of the medium-specific part of the decomposition \(\overset{\text{med}}{a}{}_k \overset{\text{med}}{p}{}_{\vec{\mu}_n}^k\) which arises solely form the medium.

(\(iv\)) The medium-specific part of the decomposition has \(k_\text{max} = i_\text{max} - j_\text{max}\) independent tensor structures and the medium-specific projection analogous to \eqref{eq:vacproj} is
\begin{align}
	\overset{\text{med}}{c}{}^{k'} & = \langle O_{\vec{\mu}_n} \rangle\; \overset{\text{med}}{p}{}^{{\vec{\mu}_n} \, k'} \nonumber\\
	& = \overset{\text{med}}{a}{}_k \overset{\text{med}}{p}{}_{\vec{\mu}_n}^k \overset{\text{med}}{p}{}^{{\vec{\mu}_n} \, k'} = \overset{\text{med}}{a}{}_k \overset{\text{med}}{\mathbf{P}}{}^{kk'}_n \, .
\end{align}
Accordingly, the final expression of the medium-specific decomposition coefficients reads
\begin{align}
	\overset{\text{med}}{a}{}_k & = \bigg(\overset{\text{med}}{\mathbf{P}}{}_n \bigg)^{-1}_{kk'} \overset{\text{med}}{c}{}^{k'} \, .
	\label{eq:medresult}
\end{align}

\section{Example: operators with four Lorentz indices}
\label{sec:ex4q}
\nin
According to Eq.~\eqref{eq:ansatzspli}, a Gibbs average carrying four Lorentz indices, \(\langle O_{\mu\nu\lambda\sigma} \rangle\), reads
\begin{align}
	\langle O_{\mu\nu\lambda\sigma} \rangle = \overset{\text{vac}}{a}{}_j \overset{\text{vac}}{p}{}_{\mu\nu\lambda\sigma}^j + \overset{\text{med}}{a}{}_k \overset{\text{med}}{p}{}_{\mu\nu\lambda\sigma}^k \, ,
	\label{eq:ansatz4LInd}
\end{align}
with the vacuum decomposition structures
\begin{align}
	\overset{\text{vac}}{p}{}_{\mu\nu\lambda\sigma}^j & =  ( g_{\mu\nu} g_{\lambda\sigma} ,\, g_{\mu\lambda} g_{\nu\sigma} ,\, g_{\mu\sigma} g_{\nu\lambda} ) \, ,
	\label{eq:vacprojvec4LInd}
\end{align}
since no structures containing \(\varepsilon_{\mu\nu\lambda\sigma}\) are considered, which are orthogonal to the ones presented here due to total anti-symmetry. The inverse to \(\overset{\text{vac}}{\mathbf{P}}{}^{jj'}_4\), then becomes
\begin{align}
	 \left( \overset{\text{vac}}{\mathbf{P}}{}_4 \right)^{-1}_{jj'} & = \frac{1}{72}
	\left[
	\begin{array}{ccc}
		 5 & -1 & -1 \\
		-1 &  5 & -1 \\
		-1 & -1 &  5
	\end{array}
	\right]
\end{align}
and
\begin{align}
	\overset{\text{vac}}{c}{}^{j'} = \left( \langle O{}_\alpha{}^\alpha{}_\beta{}^\beta \rangle ,\, \langle O{}_{\alpha\beta}{}^{\alpha\beta} \rangle ,\, \langle O{}_{\alpha\beta}{}^{\beta\alpha} \rangle \right) \, .
\end{align}

Orthogonality of the medium-specific decomposition structures to the vacuum decomposition structures \eqref{eq:ortho} implies
\begin{align}
	0  & = \left(  A_i P^i_{\mu\nu\lambda\sigma} - \overset{\text{vac}}{a}{}_{j'} \overset{\text{vac}}{p}{}_{\mu\nu\lambda\sigma}^{j'} \right) \overset{\text{vac}}{p}{}^{{\mu\nu\lambda\sigma}\, j}
	\label{eq:ortho4LInd}
\end{align}
with
\begin{align}
	P^i_{\mu\nu\lambda\sigma} = &\, ( g_{\mu\nu}g_{\lambda\sigma},\, g_{\mu\lambda}g_{\nu\sigma},\, g_{\mu\sigma}g_{\nu\lambda}, \nonumber\\[1mm]
		&\; g_{\mu\nu} v_\lambda v_\sigma,\, g_{\mu\lambda} v_\nu v_\sigma,\, g_{\mu\sigma} v_\nu v_\lambda, \nonumber\\
		&\; g_{\lambda\sigma}	v_\mu v_\nu,\, g_{\nu\sigma} v_\mu v_\lambda,\, g_{\nu\lambda} v_\mu v_\sigma, \nonumber\\[1mm]
		&\;	v_\mu v_\nu v_\lambda v_\sigma) \, .
		\label{eq:projvec4LInd}
\end{align}
The linear system of equations \eqref{eq:ortho4LInd} yields the following relations among the coefficients \(A_i\) and \(\overset{\text{vac}}{a}{}_j\):
\begin{align}
	A_j & = \overset{\text{vac}}{a}{}_j - \frac{v^2}{4}\left( A_{j+3} + A_{j+6} + \frac{v^2}{6} A_{10} \right) 
	\label{eq:3rel} \, .
\end{align}
with \(j = 1\), \(2\) and \(3\). Employing the relations \eqref{eq:3rel} to Eq.~\eqref{eq:ansatzfullmed} gives
an expression, where the vacuum decomposition can be read off easily (cf.\ Eqs.~\eqref{eq:ansatz4LInd} and \eqref{eq:vacprojvec4LInd}) and the medium-specific decomposition structures can be identified according to \eqref{eq:ansatz4LInd} as 
\begin{align}
	\overset{\text{med}}{p}{}^k_{\mu\nu\lambda\sigma} = \bigg( & g_{\mu\nu} g_{\lambda\sigma} - 4 g_{\mu\nu} \frac{v_\lambda v_\sigma}{v^2} ,\,  g_{\mu\lambda} g_{\nu\sigma} - 4 g_{\mu\lambda} \frac{v_\nu v_\sigma}{v^2} , \nonumber\\
	&  g_{\mu\sigma} g_{\nu\lambda} - 4 g_{\mu\sigma} \frac{v_\nu v_\lambda}{v^2}  ,\,  g_{\mu\nu} g_{\lambda\sigma} - 4 g_{\lambda\sigma} \frac{v_\mu v_\nu}{v^2} , \nonumber\\
	&  g_{\mu\lambda} g_{\nu\sigma} - 4 g_{\nu\sigma} \frac{v_\mu v_\lambda}{v^2}  ,\,  g_{\mu\sigma} g_{\nu\lambda} - 4 g_{\nu\lambda} \frac{v_\mu v_\sigma}{v^2} , \nonumber\\
	&  g_{\mu\nu} g_{\lambda\sigma}  + g_{\mu\lambda} g_{\nu\sigma} + g_{\mu\sigma} g_{\nu\lambda} - 24 \frac{v_\mu v_\nu v_\lambda v_\sigma}{v^4} \bigg)
	\label{eq:medprojvec4LInd}
\end{align}
with corresponding coefficients
\begin{align}
	\overset{\text{med}}{a}{}_k & = -\frac{v^2}{4} A_{k+3} \quad\text{for}\quad k = 1,\ldots,6 \, ,\\
	\overset{\text{med}}{a}{}_7 & = -\frac{v^4}{24} A_{10} \, .
\end{align}
Equipped with the medium-specific decomposition structures \(\overset{\text{med}}{p}{}^k_{\mu\nu\lambda\sigma}\) we use
\begin{align}
	& \bigg( \overset{\text{med}}{\mathbf{P}}{}_4  \bigg)^{-1}_{kk'} = \nonumber\\
	& \;\; \frac{1}{240}
	\left[
	\begin{array}{ccccccc}
		           7 & -\frac{1}{2} & -\frac{1}{2} &            2 & -\frac{1}{2} & -\frac{1}{2} &          -1 \\
		-\frac{1}{2} &            7 & -\frac{1}{2} & -\frac{1}{2} &            2 & -\frac{1}{2} &          -1 \\
		-\frac{1}{2} & -\frac{1}{2} &            7 & -\frac{1}{2} & -\frac{1}{2} &            2 &          -1 \\
		           2 & -\frac{1}{2} & -\frac{1}{2} &            7 & -\frac{1}{2} & -\frac{1}{2} &          -1 \\
		-\frac{1}{2} &            2 & -\frac{1}{2} & -\frac{1}{2} &            7 & -\frac{1}{2} &          -1 \\
		-\frac{1}{2} & -\frac{1}{2} &            2 & -\frac{1}{2} & -\frac{1}{2} &            7 & 				 -1 \\
		          -1 &           -1 &           -1 &           -1 &           -1 &           -1 & \frac{4}{3} 
	\end{array}
	\right]
\end{align}
and
\begin{align}
	& \overset{\text{med}}{c}{}^{k'}  = \nonumber\\
	&\;\;\bigg(
	  \langle O{}_\alpha{}^\alpha{}_\beta{}^\beta - 4 O{}_\alpha{}^\alpha{}_{\beta\gamma} \frac{v^\beta v^\gamma}{v^2} \rangle ,\,
		\langle O{}_{\alpha\beta}{}^{\alpha\beta} - 4 O{}_{\alpha}{}_\beta{}^{\alpha}{}_\gamma \frac{v^\beta v^\gamma}{v^2} \rangle , \nonumber\\
	&\;\;\phantom{\bigg(}
		\langle O{}_{\alpha\beta}{}^{\beta\alpha} - 4 O{}_{\alpha\beta\gamma}{}^\alpha \frac{v^\beta v^\gamma}{v^2} \rangle ,\,
		\langle O{}_\alpha{}^\alpha{}_\beta{}^\beta  - 4 O{}_{\alpha\beta\gamma}{}^\gamma \frac{v^\alpha v^\beta}{v^2} \rangle , \nonumber\\
	&\;\;\phantom{\bigg(}
		\langle O{}_{\alpha\beta}{}^{\alpha\beta} - 4 O{}_{\alpha\beta\gamma}{}^\beta \frac{v^\alpha v^\gamma}{v^2} \rangle ,\,
		\langle O{}_{\alpha\beta}{}^{\beta\alpha} - 4 O{}_{\alpha\beta}{}^\beta{}_\gamma \frac{v^\alpha v^\gamma}{v^2} \rangle , \nonumber\\
	&\;\;\phantom{\bigg(}
		\langle O{}_\alpha{}^\alpha{}_\beta{}^\beta  + O{}_{\alpha\beta}{}^{\alpha\beta} + O{}_{\alpha\beta}{}^{\beta\alpha} - 24 O_{\alpha\beta\gamma\delta}\frac{v^\alpha v^\beta v^\gamma v^\delta}{v^4} \rangle
	\bigg)
	\label{eq:medcond4LInd}
\end{align}
to arrive at the desired medium-specific decomposition coefficients \eqref{eq:medresult}. We emphasize that the medium-specific part of the decomposition \(\overset{\text{med}}{a}{}_k \overset{\text{med}}{p}{}^k_{\mu\nu\lambda\sigma}\) corresponds to the symmetric and traceless part of the Gibbs averaged operator \(\langle \mathrm{ST} O_{\mu\nu\lambda\sigma} \rangle\) with respect to Lorentz indices (cf.\ higher-twist operators in \(\rho\) meson OPE \cite{hatsukoike,leupold98}).

The condensates listed in \eqref{eq:medcond4LInd} illustrate the typical structure of medium-specific condensates composed of both vacuum and non-vacuum structures.

\section{Numerical example for a selected four-quark term}
\label{sec:numex}
\nin
Extending the in-medium OPE of pseudo-scalar \(D\) and \(B\) mesons (cf.\ \cite{hilger09} for some details) by four-quark condensate contributions to mass dimension 6 yields the Gibbs average \(\langle \bar{q} \gamma_\mu \left[ D_\nu , G_{\lambda\sigma} \right] q \rangle\) contracted with combinations of the meson's four-momenta \(q^\mu\) and the metric tensor \(g^{\mu\nu}\). 
Employing the decomposition as exercised in Sec.~\ref{sec:ex4q} its contribution to the OPE %of pseudo-scalar \(D\) and \(B\) meson
ater Borel transformation reads
\begin{align}
	 \widehat{\Pi}(M) & = \frac{1}{3} \frac{e^{-m_Q^2/M^2}}{M^2} \left( 1 + \frac{1}{2} \frac{m_Q^2}{M^2} \right) g \overset{\text{vac}}{c} \nonumber\\
	& + \frac{1}{6} \frac{e^{-m_Q^2/M^2}}{M^2} \left( 1 - \frac{1}{2} \frac{m_Q^2}{M^2} \right) g \left( \overset{\text{med}}{c}{}_1 + \overset{\text{med}}{c}{}_2 \right) \, .
\end{align}
with
\begin{align}
	\overset{\text{vac}}{c} & = g \langle \bar{q} \gamma^\mu t^A q \sum_f \bar{q}_f \gamma_\mu t^A q_f \rangle \, , \nonumber\\
	\overset{\text{med}}{c}{}_1 & = \langle g \bar{q} \gamma^\mu t^A q \sum_f \bar{q}_f \gamma_\mu t^A q_f - \frac{4}{v^2} \bar{q} \gamma^\mu \left[ (vD) , G_{\mu\nu} \right] v^\nu q\rangle \, , \nonumber\\
	\overset{\text{med}}{c}{}_2 & = g \langle \bar{q} \gamma^\mu t^A q \sum_f \bar{q}_f \gamma_\mu t^A q_f - \frac{4}{v^2} \bar{q} \slashed{v} t^A q \sum_f \bar{q}_f \slashed{v} t^A q_f \rangle \, ,
\end{align}
where the last two lines denote medium-specific condensates which vanish in vacuum.

Due to the common problem of poorly known numerical values of four-quark condensates we use (for the purpose of demonstration only) the factorization ansatz \cite{jin}. The Gibbs average of the resulting two-quark condensates at low temperatures in linear nuclear density (\(\rho\)) approximation leads to
\begin{align}
	\langle \bar{q} \gamma^\mu t^A q \sum_f \bar{q}_f \gamma_\mu t^A q_f \rangle & = -\frac{4}{9} \kappa_1(\rho) \langle \bar{q}q \rangle_0^2 \left( 1 - \frac{2\sigma_N}{m_\pi^2 f_\pi^2} \rho \right) \, , \\
	\langle \bar{q} \slashed{v} t^A q \sum_f \bar{q}_f \slashed{v} t^A q_f \rangle/v^2 & = -\frac{1}{9} \kappa_2(\rho) \langle \bar{q}q \rangle_0^2 \left( 1 - \frac{2\sigma_N}{m_\pi^2 f_\pi^2} \rho \right) \, ,
\end{align}
where the quantities \(\kappa_{1,2}(\rho)\) parameterize deviations from the naively factorized result and may be density dependent, which requires \(\kappa_1(0) = \kappa_2(0)\) to ensure vanishing of the medium-specific condensate \(\overset{\text{med}}{c}{}_2\) in vacuum. For the sake of simplicity we assume further (\(i\)) \(\kappa_1(\rho) = \kappa_2(\rho) = 1\) and (\(ii\)) the density dependence of \(\overset{\text{med}}{c}{}_1\) is dominated by the four-quark condensate part.
\begin{figure}[t] 
\centerline{\includegraphics[width=6cm]{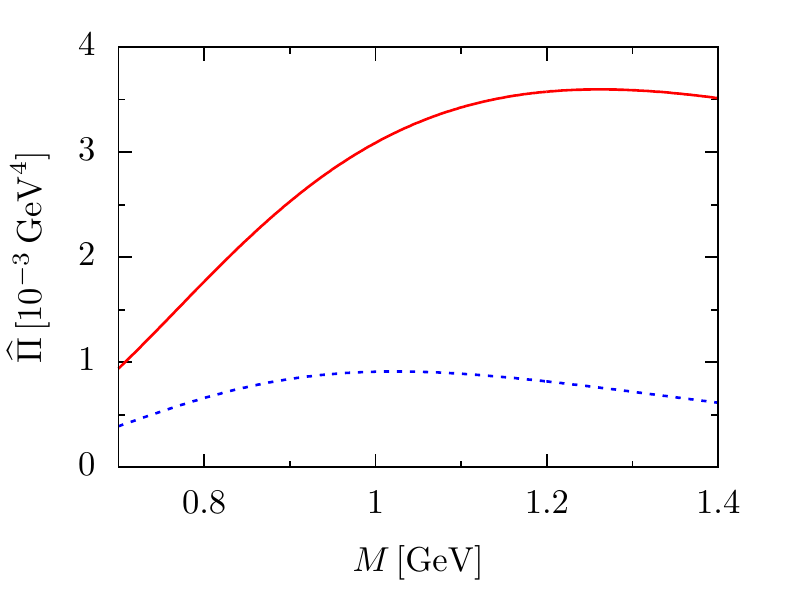}}
%{\epsfig{figure=mpsi2mc.eps,height=70mm}}
\caption{\scriptsize Borel curves of the individual pseudo-scalar D meson OPE contribution containing \(\langle \bar{q} \gamma_\mu \left[ D_\nu , G_{\lambda\sigma} \right] q \rangle\) in vacuum (red solid curve) and at saturation density (dotted blue curve, zero temperature).}
\label{fig1} 
\end{figure} 

We employed the numerical parameters from Ref.~\cite{hilger09} for \(D\) mesons to obtain the Borel curves in Fig.~\ref{fig1} exhibiting a significant difference of the contribution in vacuum and at saturation density in the Borel window around 1\;GeV dictated by stability criteria of the \(D\) meson sum rule evaluation using condensates up to mass dimension~5.

\section{Conclusions}
\nin
Motivated by large-scale experiments at FAIR, aimed at investigating properties of charm (bottom) mesons in a strongly interacting environment, we study in-medium QCD sum rules for \(D\) and \(B\) mesons. The corresponding OPE up to and including mass dimension 6 condensates requires a suitable decomposition of operators with uncontracted Lorentz indices ensuring the separation of vacuum and medium-specific contributions. The medium-specific condensates appear as coefficients vanishing, by definition, in vacuum. A specific example is provided for the splitting into vacuum and medium-specific four-quark condensates. Such four-quark condensates can qualify as order parameters of chiral restoration \cite{hilger12,hohlerrapp14}, which would be masked otherwise. Thus, the avenue is prepared for elaborating the QCD sum rule in an analog manner as exercised previously for the \(\rho\) meson, where four-quark condensates play a decisive role.

\section*{Acknowledgements}
\nin
This work is support by BMBF 05P12CRGHE and by the Austrian Science Fund (FWF) under project no. P25121-N27.

%%

% References with bibTeX database:

\bibliographystyle{model1a-num-names}
\bibliography{QCD14_lit}
% Authors are advised to submit their bibtex database files. They are
% requested to list a bibtex style file in the manuscript if they do
% not want to use elsarticle-num.bst.





% \end{thebibliography}

%%%%%%%%%%%%%%%%%%%%
%\vfill\eject

%\input{bib_sample}

\end{document}